\documentclass{aa}
\usepackage{graphicx}
\usepackage{amssymb}

\newcommand{\fu}{erg~cm$^{-2}$~s$^{-1}$}
\def\sax{{\it Beppo}SAX}
\def\grb{GRB 020410}
\def\nh{\ifmmode{N_{\rm H}}\else{$N_{\rm H}$}\fi}
\def\ltsima{$\; \buildrel < \over \sim \;$}
\def\simlt{\lower.5ex\hbox{\ltsima}} 
\def\gtsima{$\; \buildrel > \over \sim \;$}
\def\simgt{\lower.5ex\hbox{\gtsima}} 

\def\lapp{\ifmmode\stackrel{<}{_{\sim}}\else$\stackrel{<}{_{\sim}}$\fi}
\def\gapp{\ifmmode\stackrel{>}{_{\sim}}\else$\stackrel{>}{_{\sim}}$\fi}

\begin{document}
\title{Multiwavelength study of the very long \grb}
\author{Luciano Nicastro\inst{1}
\and J. J. M. in 't Zand\inst{2,3}
\and L. Amati\inst{6}
\and E. Mazets\inst{4}
\and A. Castro-Tirado\inst{5}
\and J. Gorosabel\inst{5}
\and D. Lazzati\inst{7}
\and E. Costa\inst{8}
\and M. De Pasquale\inst{8}
\and M. Feroci\inst{8}
\and F. Frontera\inst{6}
\and J. Heise\inst{2,3}
\and E. Pian\inst{9}
\and L. Piro\inst{8}
\and C. S\'anchez-Fern\'andez\inst{10}
\and P. Tristram\inst{11}
}
\institute{
 IASF--CNR, Via U. La Malfa 153, 90146 Palermo (I)
\and
 SRON National Institute for Space Research, Sorbonnelaan 2, 3584 CA Utrecht,
 The Netherlands
\and
 Astronomical Institute, Utrecht University, PO Box 80000, 3508 TA Utrecht,
 The Netherlands
\and
 Ioffe Institute, 26 Polytekhnicheskaya, St. Petersburg 194021,
 Russian Federation 
\and
Instituto de Astrof\'{\i}sica de Andaluc\'{\i}a (IAA-CSIC),
P.O. Box 03004, E-18080 Granada, Spain
\and
 IASF--CNR, Via P. Gobetti 101, 40129 Bologna (I)
\and
 Institute of Astronomy, University of Cambridge, Madingley Road,
 UK CB3 0HA Cambridge, UK
\and
 IASF--CNR, Via Fosso del Cavaliere, 00131 Roma (I)
\and
 Osservatorio Astr. di Trieste, Via G.B. Tiepolo 11, 34131 Trieste (I)
\and
 ESA-VILSPA, Villafranca del Castillo, PO Box 50727, 28080 Madrid, Spain
\and
 Mt. John University Observatory, Canterbury University, New Zealand
}

\offprints{L. Nicastro, \email{nicastro@pa.iasf.cnr.it}}
\date{Received 24 March 2004 / Accepted }

\abstract{
GRB 020410 is by far the longest $\gamma$-ray burst (with a duration
of about 1600~s) to have been followed up from the X-ray through the
radio regime.  Afterglow emission was detected in X-rays and at
optical wavelengths whereas no emission was detected at 8 GHz brighter
than 120 $\mu$Jy.  The decaying X-ray afterglow, back extrapolated to
11 hours after the burst, had a flux of $7.9\times10^{-12}$ \fu\ (2--10
keV); the brightest detected so far. No direct redshift determination
is available yet for this GRB, but according to the empirical
relationship between the peak energy in the $\nu F_\nu$ spectrum and
the isotropic energy output, $z$ is constrained in the range 0.9--1.5.
The reconstructed optical afterglow light curve implies at least two
breaks in the simple power-law decay.  This may be related to
emergence of a SN, or refreshment of the external shock by a variation
in the circumstellar medium. By comparing the backward extrapolation
of the 2--10 keV afterglow decay, it is shown that the long duration of
the prompt emission is not related to an early onset of afterglow
emission, but must be related to prolonged activity of the ``central
engine''.}

\maketitle              

\section{Introduction}

Gamma-ray Bursts (GRB) show great diversity with regard to both their
durations and spectral properties. GRBs last from a fraction of a
second to thousands of seconds, as established by the BATSE survey
(e.g. \cite{Paciesas99}). Prompt X-ray counterparts of GRBs, detected
by Ginga, \sax, HETE-2 have a very wide distribution of intensities
and durations.  Tails and precursors of X-ray counterparts were also
observed by WATCH/Granat (\cite{ct94}).  Those events characterized by
an X-rays-to-$\gamma$-rays (2--10/40--700 kev) fluence ratio larger
than $\sim0.5$ are classified as X--ray rich (e.g. \cite{Feroci01}).
Moreover, transient X--ray sources with characteristics similar to
those of GRB counterparts, although with no simultaneous GRB detection
(so called ``X-ray flashes''; \cite{Heise01, moc03}) were detected by
the \sax\ Wide Field Cameras (WFC) and, subsequently, by the
HETE--2/FREGATE instrument. Recently (\cite{Zand03}), have reported
the detection of 4 long, faint X-ray transients during sky surveys
with the \sax-WFC. Three of these are confirmed GRBs, because they
coincide with BATSE detections. They show durations ranging from 540 s
to 2550 s and are characterized by a mildly soft spectrum.

The very different ratios of $\gamma$-ray vs X-ray peak fluxes or
fluences point either to different viewing angles of the relativistic
jets in which GRBs are formed (e.g. \cite{Granot02, Yamazaki02}) or to
a different amount of baryon contamination of the fireball
(e.g. \cite{Dermer99, Huang02}). Furthermore, the existence of a class
of GRBs with long X-ray durations is important for the investigation
of the connection between the prompt and afterglow components and may
also suggest a high redshift origin.  However, redshift constraints
imposed on XRF 020903, 030723 and, possibly, 031203 do not support the
high redshift scenario (\cite{Soderberg03, Prochaska03, Fynbo04}).

\grb, first detected in X-rays only by the \sax-WFC
(\cite{Gandolfi02}), stands out for its long duration, more than 1500
s in the 2--28 keV band (see Sect. \ref{wfc_an}), and for the relative
weakness of its $\gamma$-ray signal, detected by Konus-Wind in an
offline analysis.  Based on its X-ray-to-$\gamma$-ray fluence ratio
(see Sect. 3), \grb\ lies in the soft tail of genuine GRBs and
marginally qualifies as an ``X-ray rich'' GRB (see \cite{Heise01}).
Upon detection of the GRB we started an X-ray and optical search and
monitoring campaign of its afterglow.  We present here the results of
our study of the prompt and afterglow emission.

\section{Observations}

The \sax\ Wide Field Camera unit 2 (\cite{jager97}) detected a
transient event on April 10, 2002, 10:39:40 UT ($T_0$) at coordinates
${\rm RA} = 22^{\rm h} 07^{\rm m} 04^{\rm s}$, ${\rm Dec} = -83\degr
49' 18''$.  The off-axis angle was 10\fdg9 in a field of view of
$40\degr \times40\degr$.  Due to battery efficiency degradation, the
\sax\ Gamma-Ray Burst Monitor (GRBM) was switched off so the
association of the WFC event with a GRB could not be immediately
verified.  After the WFC detection, lacking a GRBM measurement, we
searched the ratemeters of other $\gamma$-ray monitors.  The most
sensitive instrument, apart from the GRBM, with low energy coverage is
Konus on the {\em Wind} satellite (Aptekar et al. 1995). An off-line
analysis of Konus data allowed us to confirm the GRB nature of the
event (see Sect. 1.3).  In order to exclude the possible
recurrent/flaring nature of the source detected by the WFC, we
performed a detailed archival analysis on all the \sax-WFC data around
the \grb\ position (covering the period 1996--2001). There were 186
observing periods with a net exposure time of 2.8 Ms (32 days).  The
2--10 keV flux limit was 0.2 mCrab or $\simeq4\times10^{-12}$ \fu\
(this flux level is similar to that measured by the \sax\ ToO
observations; see below).  No stable or flaring source was detected at
the position of the candidate GRB.  The 2\arcmin-radius GRB error
circle determined by the WFC was disseminated 4 hours after the event
(\cite{Gandolfi02}).  This prompted a fast pointing at the error box
of the \sax\ Narrow Field Instruments (NFIs), aimed at the detection
of the early X-ray afterglow (Section 2), and rapid searches for the
optical (Section 3) and radio counterpart from the Southern
hemisphere.  The follow-up with the \sax-MECS (see Section 2) resulted
in a reduction of the error region; an error box of $20\arcsec$ radius
was disseminated on April 18. A subsequent search for a radio
counterpart at 8.47~GHz on April 23 yielded an upper limit of
200~$\mu$Jy (\cite{frail02}).

\section{Prompt emission}
\subsection{WFC data analysis}\label{wfc_an}

\begin{figure}[ht]
\centerline{\includegraphics[width=\columnwidth, clip=]{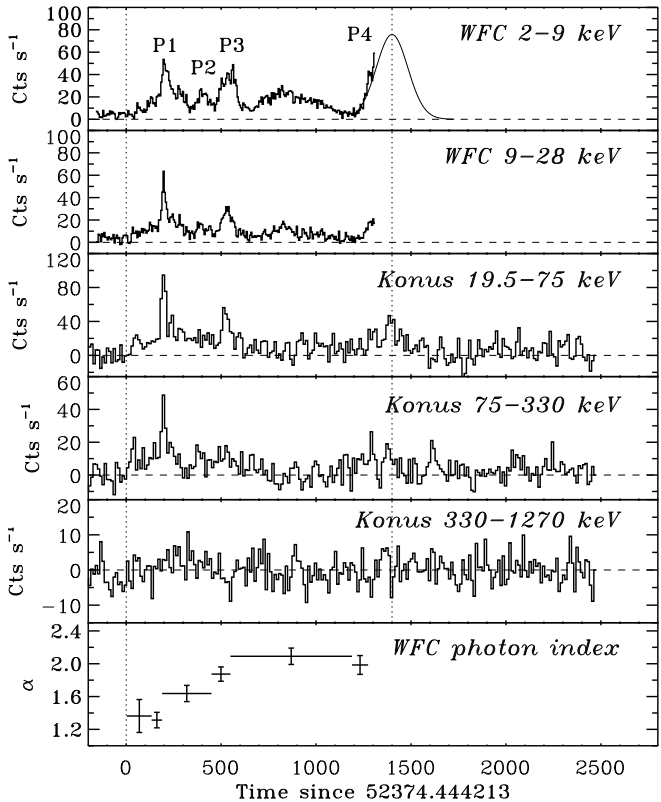}}
\caption{ Upper 5 panels: photon count rates in two WFC bands and 3
Konus bands.  A time interval of 1.48~s was subtracted from the Konus
times to account for the delay in light travel time between {\em Wind}
and \sax.  The extrapolated (Gaussian) last peak is shown in the 2--9
keV light curve.  The lowest panel shows the photon index resulting
from a power-law fit to the WFC spectra only.}
\label{f1_promptlc}
\end{figure}

The WFC light curves of \grb\ are presented in Fig.~\ref{f1_promptlc}.
The data are background subtracted using rates of 29 and 28 cts
s$^{-1}$ for the lower and higher energy bands respectively. These
values were determined from the light curve portions immediately
preceding the event and are appropriate for the part of the detector
illuminated by the source.  Note that, due to the slow rise of the GRB
flux at its onset, we set the start time by interpolating the rise of
the first peak with an exponential function; $T_0$ corresponds to the
time when the function reaches the zero counts level.  Although this
method is somewhat arbitrary, it is reliable, because our results do
not change significantly if we move forward the start time by one
minute.  In the 16.1 hours time interval prior to the event, no signal
was detected.  The first sign of activity is distinguishable at high
energies about 15 s before the adopted start time.  The event consists
of four main pulses (hereafter designated as P1, P2, P3 and P4), the
first of which, P1, has a maximum peak flux of
$(1.9\pm0.4)\times10^{-8}$ \fu\ in the 2--10 keV range. Earth
occultation caused the end of the observation when the GRB was still
active and rising, therefore we can only put a lower limit on its
duration and fluence, $T_d \simgt 1300$ s and
$4.7\times10^{-6}$~erg~cm$^{-2}$ (2--10 keV), respectively.  The
truncated peak (P4) has a 2--10 keV flux $F_{\rm
p}=(2.5\pm0.4)\times10^{-8}$ \fu\ and it is then the highest peak in
this range.

However we can make reasonable assumptions to estimate the missing
flux due to the truncated peak:
\begin{itemize}
\item its shape is Gaussian, i.e. symmetric (usually the decay is
  slower than the rise so this gives a lower limit to the pulse
  width);

\item it is aligned with the Konus (G1) one (see below);

\item its FWHM conforms to the law ${\rm FWHM}\propto E^\alpha$ (see below);

\item the total counts are proportional to that of P3 when
  compared to the Konus (G1) one, i.e. the spectral shape is the same
  as for P3.
\end{itemize}
The extrapolated light curve is shown in Fig.~\ref{f1_promptlc}.  We
find then that, including the reconstructed portion of the light
curve, the event lasts $\sim 1550$ s; the total counts are increased
with respect to the measured ones by 22\% in the energy band 2--9 keV
and by 21\% in 9--28 keV.  The estimated fluences in 2--10 and 2--28
keV become $(5.7\pm0.5)\times10^{-6}$ and $(1.1\pm0.1)\times10^{-5}$
erg~cm$^{-2}$ respectively.

With its large X-ray fluence, \grb\ ranks in the top 3.4\% of the
\sax-WFC GRB+XRFs sample.

The temporally resolved WFC spectra are described by simple power-laws
$N_E\propto E^{-\alpha}$ with $\alpha$ varying between $\simeq 1.4$ at
start to $\simeq 2.1$ from about $T_0 + 700$ s onward (see
Fig.~\ref{f1_promptlc}, bottom panel).  This index range is typical
for GRBs (\cite{Preece00, Frontera00}), even though on the soft side
of the distribution and its change indicates a hard--to--soft
evolution. No evidence of intrinsic absorption is found in any of the
spectra; the average $2\sigma$ upper limit on the intrinsic hydrogen
equivalent column density is $3.3\times10^{21}$~cm$^{-2}$ .

\subsection{Konus data analysis}

Since the event did not trigger Konus in a burst data acquisition
mode, we analyzed the non-triggered data, pertaining to photon count
rates at 2.944~s resolution in 3 bands: 19.5--75 keV (G1 band),
75--330 keV (G2) and 330--1270 keV (G3).  The corresponding time light
curves are shown in Fig.~\ref{f1_promptlc}.  The background was
estimated in an analogous way as for the WFC, giving 2923 (G1), 1014 (G2)
and 420 cts~s$^{-1}$ (G3).  These are close to the nominal
values, which have an accuracy of 2\%.  The event is clearly
detected by Konus up to a few hundred keV. This confirms the GRB
nature of the event detected by the \sax-WFC. There appears to be an
initial signal in the G2 band at about the same time as in the upper
WFC band, and the main Konus peak coincides with the first WFC
peak. The other WFC peaks are also detected with Konus. A fortunate
circumstance is that Konus continued to observe after the WFC pointing
ended. The Konus data in the G1 band suggest that the X-ray flux
continues to increase for about 100~s past the end of the WFC
observation. Therefore, the absolute flux maximum in X-rays was most
likely missed.

The spectra obtained by combining the data in the 3 Konus channels are
not described by a simple power-law.  The flux in the G3 band must be
slightly suppressed to satisfy the observed count rate. If this is
modeled by a power-law with an exponential cut off, we find that a
shallow e-folding cutoff energy of 900~keV, in combination with a
photon index of 1.8, is sufficient to explain this suppression.  WFC
and Konus are not inter-calibrated instruments. Fortunately they have
an overlapping energy band so to get the flux in the Konus band we
allowed for a free normalization respect to the WFC in a joint fit
(keeping constant photon index and cut off energy).  The average
photon spectrum of WFC and Konus data may be described by the law
$1.0 E^{-1.8}{\rm exp}(-E/900)$~ph~cm$^{-2}$s$^{-1}$~keV$^{-1}$.
The 15--1000 keV fluence is $2.8\times10^{-5}$
erg~cm$^{-2}$ (for 1550 s) and the peak flux $1.0\times10^{-7}$ \fu.
The 2--10 keV to 15--1000 keV fluence and peak flux ratios are then
$\sim 0.20$ and $\sim 0.25$ respectively.  The fluence in the standard
BATSE energy range, 50--300 keV, is $\sim 1.3\times10^{-5}$
erg~cm$^{-2}$, so that in this case the soft to hard fluence ratio is
$\sim 0.44$.  If we consider the \sax-GRBM energy band 40--700 keV
instead, the fluence is $\sim 2.1\times10^{-5}$ erg~cm$^{-2}$, giving
a ratio of $\sim 0.27$.

\begin{figure}
\centerline{\includegraphics[width=\columnwidth]{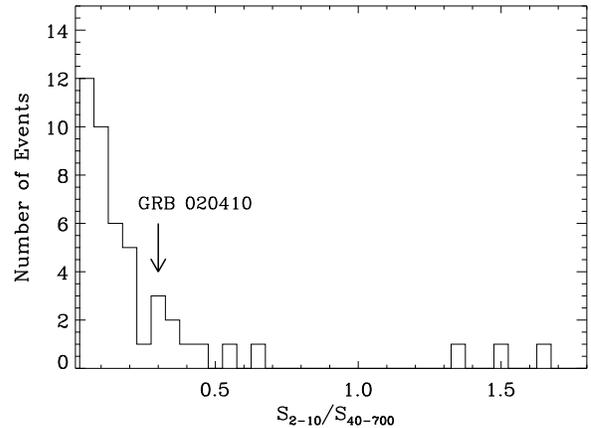}}
\caption{Histogram of 2--10/40--700 keV fluence ratio for the \sax\
GRBs.}
\label{f2_fratio}
\end{figure}

Though these values are not particularly high, they put \grb\ in
between typical GRBs and X-ray rich events (see Fig.~\ref{f2_fratio},
\cite{Feroci01} and Fig. 3 in \cite{Heise01}).
Finally, by integrating the background subtracted Konus light curves from
$t=1550$ to $t=2500$ s, we detect a marginally significant counts excess
of $\sim3\sigma$ for the G1 and G2 bands. No excess is found in G3.

\section{\sax\ MECS observations}

At the epoch of GRB020410, only the MECS, among the \sax-NFIs, was
operational because of the critical status of the satellite batteries
at the end of its lifetime.  The error box of the event was observed
with the MECS in Target-of-Opportunity (ToO) mode on 11 April 2002,
06:53:21 UT and on 12 April 2002, 16:48:44 UT, i.e. 20.2 and 54.3
hours after the burst onset, respectively.  The observations lasted
7.5 and 5.3 hours with exposure times of $\sim 23$ and $\sim 15$ ks,
respectively.  A third pointing was attempted but it failed because of
problems in the satellite attitude control system.

A relatively bright source was detected within the WFC error circle at
coordinates ${\rm RA} = 22^{\rm h}\, 06^{\rm m}\, 25\fs8$, ${\rm Dec}
= -83\degr\, 49'\, 27''$ (1SAX J2206.4$-$8349).  The source showed a
clear fading between the two epochs.  We therefore identify it with
X-ray afterglow emission. Thanks to the new calibration of the MECS
instrument (\cite{pc02})
we were able to set a 90\% confidence error circle of $20''$ radius.
This position was $\simeq 1\farcm3$ away from the center of the WFC error
circle (\cite{Gandolfi02}).
\begin{figure*}
\centerline{\includegraphics[width=\textwidth]{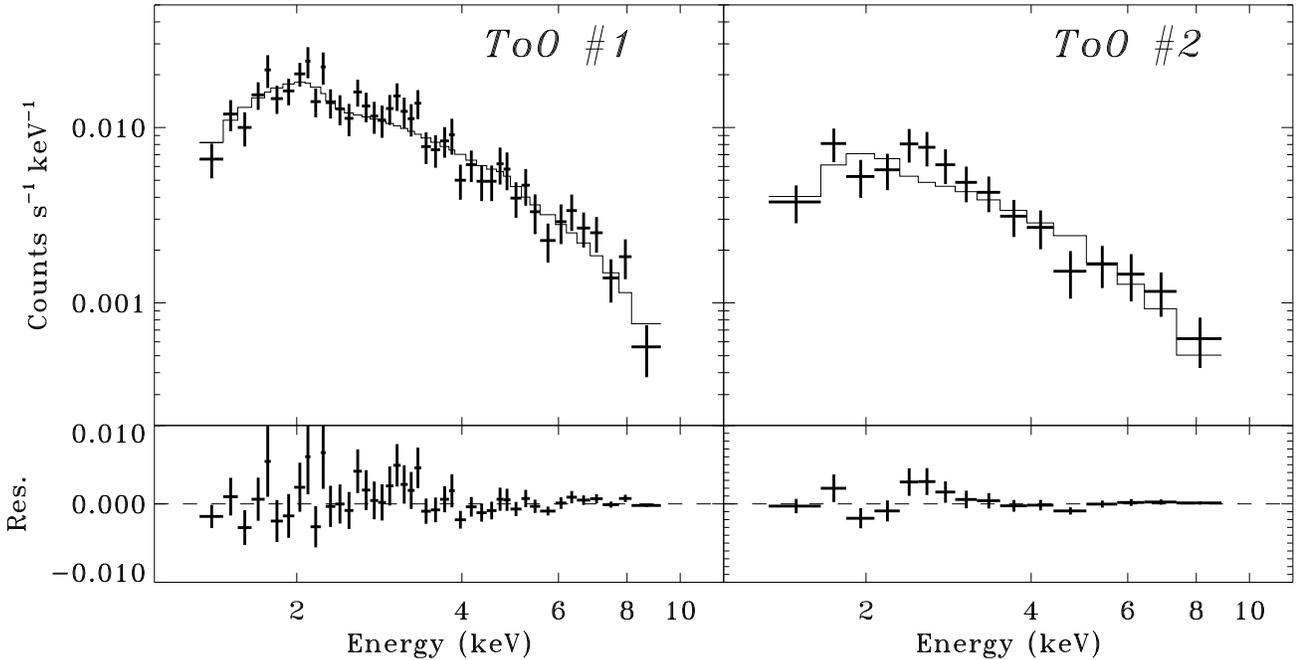}}
\caption{Spectra for the two \sax-MECS ToOs. The data are fitted with
a power-law plus a Galactic \nh. Note the excess around 3 keV in both ToOs.}
\label{f3_spe12}
\end{figure*}

\subsection{Spectral analysis}

By fitting the MECS spectra of both ToOs with single power-laws we
obtain acceptable results.  The photon index is quite stable around
$\alpha \simeq 2$ ($N_E = K E^{-\alpha}$), typical for GRBs.  The
2--10 keV average flux is $(3.4\pm 0.2)\times 10^{-12}$
and $(1.5\pm 0.1)\times 10^{-12}$ \fu\ for the first and second ToO,
respectively.  Due to the lack of the LECS data we included in our
analysis the lower energy MECS data down to 1.3 keV (instead of the
usual 1.6 keV).  We verified the validity of the response matrix and
data quality in this range.  No extra absorption above the Galactic
one is detected.  However, driven by systematic deviations in the
residuals, we also tried to fit additional components. In particular,
we added a Gaussian component to model the bump visible around 3 keV
(see Fig.~\ref{f3_spe12}).  In the first ToO the line position is
$3.07^{+0.15}_{-0.17}$ keV ($2\sigma$ confidence). The F-test chance
probability of fit improvement is 1.7\% ($2.1\sigma$).  In the second
ToO the line position is $2.56^{+0.19}_{-0.15}$ keV with a F-test
chance probability of 3.5\% ($1.8\sigma$).  Assuming the line
detection is authentic, these results suggest its energy is changing,
although a blend of multiple components could mimic this behavior
instead (see below).  Next, we fitted the first ToO split in two,
both with a simple power-law and with an additional Gaussian.  Also in
this case the line position, though at $2\sigma$ level, is not stable,
being $2.96^{+0.18}_{-0.28}$ keV in the first and
$3.36^{+0.20}_{-0.43}$ keV in the second half of the ToO. The F-test
chance probabilities for the two sections are 5.8\% and 4.6\%,
respectively.  In Table 1 a summary for all these fits is presented.
Because the statistics are poor to confirm the presence of emission
lines, our confidence level about the above reported results are (as
shown) $2\sigma$ level results.  We performed 3000 spectral
simulations (1000 for each of the halves of ToO~1 and 1000 for ToO~2)
of a power-law with parameters fixed to those obtained for our data
and checking the $\chi^2$ improvement adding a Gaussian line of width
0.1 keV falling either in the whole MECS bandpass 2--9 keV (where we
detect signal) or 2.2--4 keV (line position).  For the latter range we
derive chance probabilities not significantly different from those
derived from the F-test on the observed data. For the wider energy
range the statistical significance is a factor 1.5 lower.  By randomly
selecting one simulated spectrum of the first half of ToO~1, one of the
second half of ToO~1 and one of ToO~2, we computed the number of cases
in which a line is detected in all the three spectra with line energy
centroids within $\sim 1.2$ keV of the observed (fitted) ones and
significance equal or greater than the observed ones. This number
turns out to be 0.2\%.
\begin{table}
\caption{Spectral analysis of \grb; the \nh\ is fixed at the Galactic value
$8\times 10^{20}$ cm$^{-2}$ and the line width at 0.1 keV.}
\begin{tabular}{cccccc}
ToO &  $K_{\rm PL}^{\rm a}$  & $\alpha$  & $E_{\rm line}$ & $K_{\rm line}^{\rm a}$ & $\chi^2_n$ \\
    &  $\times 10^{-3}$  &  & keV & $\times 10^{-5}$ & \\
\hline
1  & $1.44^{+0.22}_{-0.20}$ & $2.05^{+0.12}_{-0.13}$ & -- &
 -- & 0.93 \\
1  & $1.38^{+0.22}_{-0.20}$ & $2.05^{+0.13}_{-0.12}$ & $3.07^{+0.15}_{-0.17}$ &
 $2.4^{+1.8}_{-1.4}$ & 0.80 \\
1a & $1.57^{+0.34}_{-0.30}$ & $2.06^{+0.17}_{-0.16}$ & -- &
 -- & 0.81 \\
1a & $1.49^{+0.35}_{-0.31}$ & $2.05^{+0.19}_{-0.18}$ & $2.96^{+0.18}_{-0.28}$ &
 $2.8^{+3.1}_{-2.1}$ & 0.69 \\
1b & $1.27^{+0.28}_{-0.25}$ & $2.01^{+0.16}_{-0.17}$ & -- &
 -- & 0.80 \\
1b & $1.24^{+0.28}_{-0.25}$ & $2.02^{+0.18}_{-0.18}$ & $3.36^{+0.20}_{-0.43}$ &
 $2.2^{+2.0}_{-1.9}$ & 0.70 \\
2  & $0.53^{+0.17}_{-0.14}$ & $1.93^{+0.24}_{-0.24}$ & -- &
 -- & 1.13 \\
2  & $0.44^{+0.18}_{-0.15}$ & $1.84^{+0.28}_{-0.28}$ & $2.56^{+0.19}_{-0.15}$ &
 $3.4^{+1.9}_{-2.2}$ & 0.75 \\
\hline
\end{tabular}
\begin{list}{}{}
\item[Note:] errors are 90\% confidence level.
\item[$^{\rm a}$] Power-law and the Gaussian normalization constants.
\end{list}
\end{table}

\subsection{Temporal analysis}
\begin{figure}
\centerline{\includegraphics[width=\columnwidth]{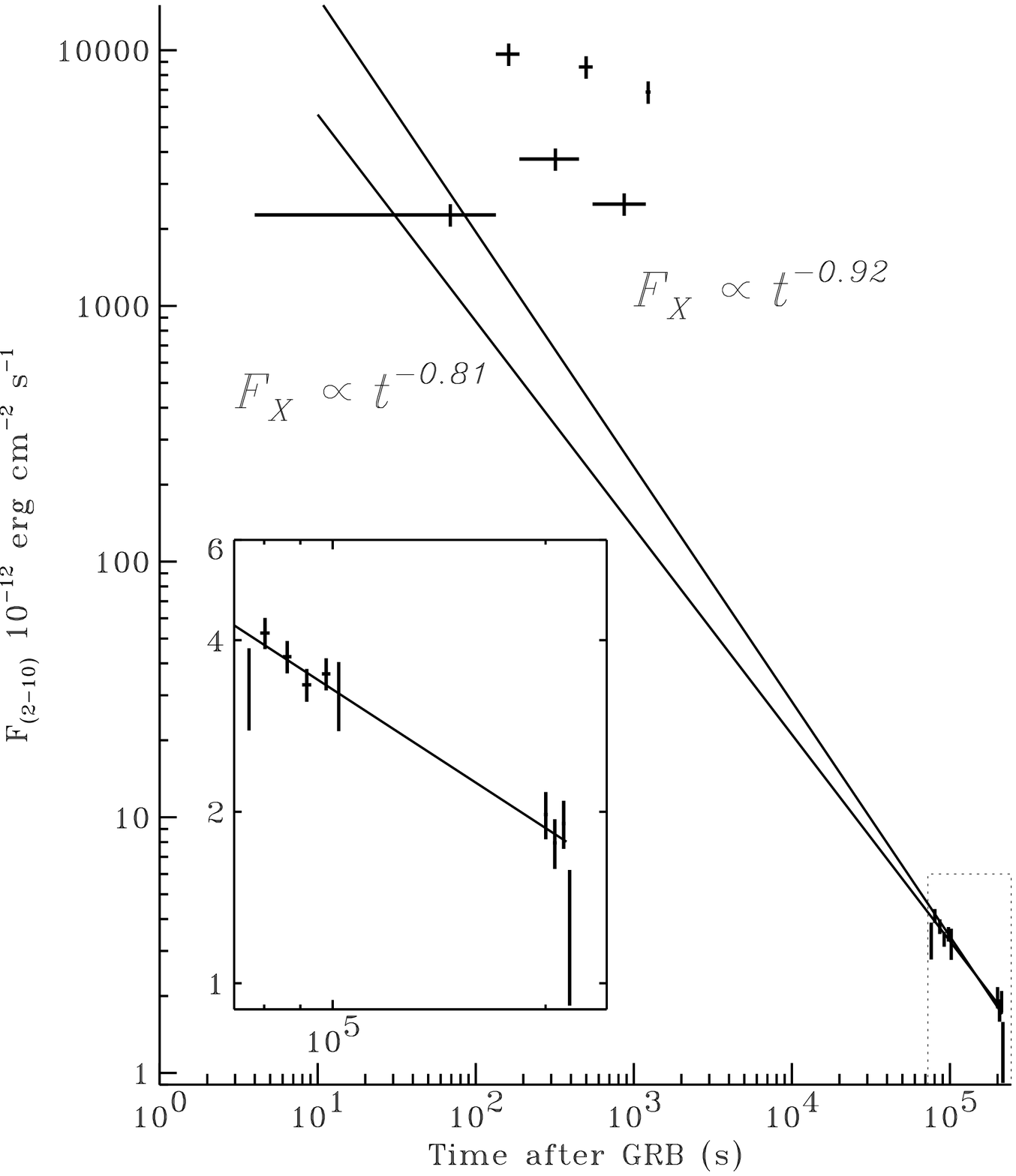}}
\caption{X-ray (2--10 keV) WFC and MECS light curve including \grb.
The flatter fit is obtained using only the MECS data; the steeper one
is instead constrained by using the 90\% upper limit to the fitted
slope.}
\label{f4_xdecay}
\end{figure}

Figure \ref{f4_xdecay} shows the MECS 2--10 keV flux measurements
together with the WFC light curve binned in the same 6 time intervals
used in Fig.~\ref{f1_promptlc} for the spectral indices. Two possible
power-law decay fits $F_x \propto t^-\delta$ are shown.  One is the
best fit model to the MECS ToO data (also shown in the inset), giving
a fairly flat decay index of $\delta = 0.81\pm 0.07$ with a reduced
$\chi^2_n$ of 0.86. The other one is the model obtained by adopting
the 90\% upper limit of the decay index (0.92).

Following De Pasquale et al. (2002), we calculated the 1.6--10 keV flux 11 hours
after the burst using the $\delta = 0.81$ decay index.  The resulting
value of $7.9\times 10^{-12}$ \fu\ is the highest among all the \sax\
detected GRBs, the second being GRB 010222 with $7.3\times 10^{-12}$
\fu\ (\cite{Zand01}).

\section{Optical observations}
\begin{figure}[ht]
\centerline{\includegraphics[width=\columnwidth]{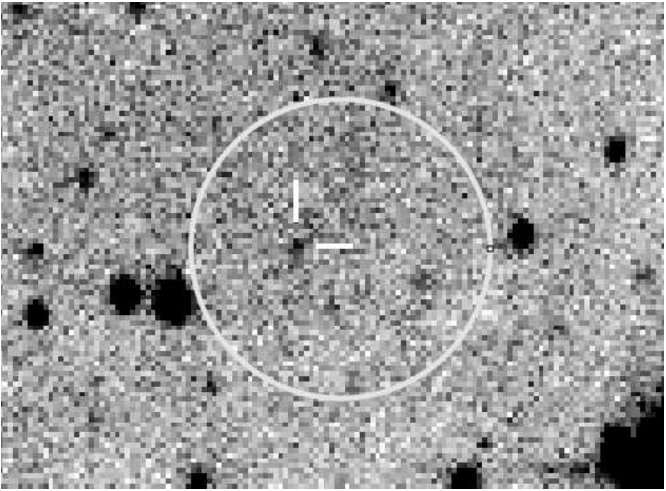}}
\caption{600 s co-added R-band image of the \grb\ field taken with the
MOA 0.6-m telescope 6.25 hours after the burst. The larger circle
represents the MECS $20''$ error box; the smaller one highlights
the OT at the HST position. Norh up and East to the left.}
\label{f5_moa}
\end{figure}

\begin{figure}
\centerline{\includegraphics[width=\columnwidth]{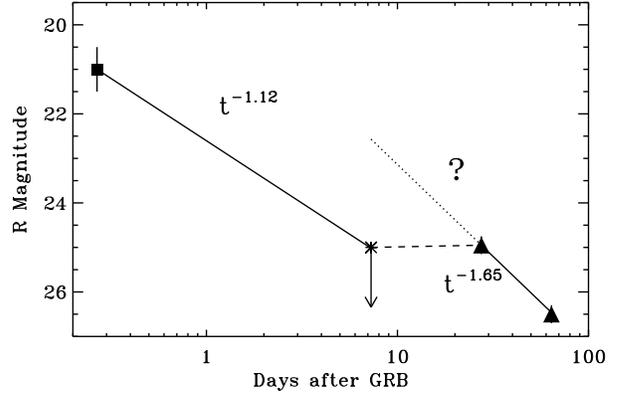}}
\caption{R-band magnitudes and possible time evolution of the optical
transient of \grb. The first point is MOA 0.6-m detection. Asterisk
is the ESO 3.6-m (see also \cite{Levan04}) upper limit. Filled
triangles are the HST measurements converted using ${\rm R}-{\rm V}\simeq 0.4$
(\cite{simon01}).
Early time decay law marginally agrees with that observed in X-rays.}
\label{f6_odecay}
\end{figure}

Two optical searches were attempted 6.0 and 6.28 hours after the GRB
onset by two groups using two telescopes of the Mount John Observatory
(MJO) located in New Zealand: the MJO 0.6-m telescope and the Microlensing
Observations in Astrophysics (MOA) 0.6-m telescope.  Several images
were collected with the two telescope for a total of 900-s
(unfiltered) for MJO 0.6-m and 800-s (wide B-band filter) and 600-s
(R-band) MOA 0.6-m.  No positive
detection was reported when compared with the POSS-II red plates down
to ${\rm R}\simeq 20$ (\cite{Kilmartin02, Albertonz02}). On April 17, images
in UBR bands were collected with the 3.6-m ESO telescope at La Silla
(\cite{Levan04}).
Further B and R bands
observations were carried out at the 8.2-m ESO VLT on June 11.
Due to the very poor seeing conditions ($2''$) and high airmass (1.95),
a limiting (R) magnitude of 24.5 for the OT is derived, which may
constrain the SN maximum brightness if an underlying SN is considered.
HST observations performed 27.5 and 64 days after
the GRB onset revealed a variable source about $10''$ away from the
MECS X-ray position (\cite{Fruchter02, Levan04}). The decay index was
$\delta \simeq 1.65$.  The V magnitudes reported, if converted to R
using the relation of $\check{\rm S}$imon et al. (2001), become 24.9 and 26.5,
respectively. Assuming a constant decay index, this would imply
${\rm R}\simeq 16.5$ 6 hours after the burst.  Driven by this finding, a
more accurate analysis of the MOA data was performed leading to the
detection of the optical transient with ${\rm R}\simeq 21$ and
${\rm B}> 21.5$ (see
Fig.~\ref{f5_moa}).  Re-analysis of the ESO 3.6-m R-band observation
performed in April 17.31--17.35 (i.e. 6.87 days after the burst), also
revealed a $2\sigma$ level source at ${\rm R}\approx 24$ compatible with
the ${\rm R}=24.3$ mag galaxy 0\farcs5 away from the OT detected by HST
(\cite{Levan04}). Then we can adopt a limit of ${\rm R}=25$ to the OT in the
ESO image so that the decay index between the MJO and ESO observation
is $\geq 1.1$.
The optical lightcurve between 10 and 20 days
must clearly deviate from this decline (see Fig.~\ref{f6_odecay}), and
this may be accounted for if the flux increases due to the emergence
of another emission component (e.g. a supernova; \cite{Woosley93}) or
a refreshed shock (\cite{Panaitescu98}).
Whichever is the correct interpretation, like for GRB 991208
(\cite{Alberto01}), a double break in the optical light curve must be invoked.
A detailed study of the OT behavior and possible explanations for the
late time re-bump is reported in Levan et al. (2004).

\section{Discussion}
\grb\ is by far the longest GRB event for which X-ray afterglow
emission and an optical counterpart (though weak) have been
discovered.  The long duration of \grb\ can be considered ``tail of
the distribution'' rather than a peculiar case, though peculiar
circumstances are required.  As the other very long GRBs detected by
the WFC (\cite{Zand03}), this event shows a high ratio between X-ray
and $\gamma$-ray fluence, although it cannot be classified as an X-ray
rich GRB. In addition, apart the high X-ray content, X-ray rich GRBs
are also characterized by the lack of optical afterglow (not true for
the X-ray flashe 030723 and possibly 020903), which is not the case
for \grb.  We note, however, that the two things could not be linked
as the rapidly fading OT of the ``not-quite'' X-ray rich GRB 021211 has
demonstrated (\cite{crew03}).
Spectroscopically, \grb\ does not show peculiarities.  This in
addition to (a) the measured upper limit on the intrinsic absorption,
(b) the ``smooth" increase of the spectral index, (c) the late time
X-ray ``afterglow" decay that does not connect to the late time prompt
emission, suggest the prompt event is all due to internal shock rather
than being a superimposition of internal and external shocks, or at
least the latter is negligible.  In the fireball model, a connection
between the long duration of the ``prompt'' event and the slow decay
of the afterglow could be accounted for e.g.  by the superimposition
of several (external) shocks produced by each of the main peak of the
internal shock, which are all relatively long when compared to normal
GRBs, causing a continuous refreshment of the external shock
(e.g. \cite{bj02}).

Unlike what is observed for several GRBs (e.g. \cite{costa99}) the
backward extrapolation of the afterglow fading law of \grb\ is
inconsistent with the flux measured during the last part of the prompt
emission (see Fig.~\ref{f4_xdecay}). This may be linked to the
extremely long duration of the event and prevents us to derive an
indication of the afterglow emission onset time.  We can estimate an
upper limit to the afterglow 2--10 keV fluence if, by following
\cite{Frontera00}, we assume that the afterglow emission starts at
63\% of the duration of the GRB and thus we integrate the fading law
between 973 s and $1\times10^6$ s.  The result is $1.96\times10^{-6}$
erg cm$^{-2}$, corresponding to about 34\% of the fluence measured in
the prompt event in the same energy range and to about 9\% of the
prompt fluence in 40--700 keV. These values are well within the
observed range of normal GRBs (\cite{Frontera00}).

Alternatively, it is possible to identify the onset of the
external shock at $t\sim500$~s when the spectrum of
the prompt emission becomes consistent with the late time MECS
spectra. In the simple case in which the fireball is homogeneous and
thin, the GRB variability should be suppressed and the lightcurve be
described as a power-law initially rising as $t^2$ and then smoothly
turning over to a decay slope which depends on the spectral range and
dynamics of the fireball (\cite{SaPi99}). In fact the lightcurve
of \grb\ is highly variable after the spectral transition,
showing a prominent emission episode at $t~1500$~s (P4 in
Fig.~\ref{f1_promptlc}).
This behavior can be understood if the inner engine does not
turn off at the end of the gamma-ray phase, but releases a sizable
amount of energy at $t\sim1500$~s. This late emission, however, should
be inefficient in the production of $\gamma$-rays or, in terms of the
internal-external shock scenario, it should avoid the internal shock phase.
The time $t\sim1500$~s is not the deceleration time of the fireball, but
the delay with which the inner engine released the fireball component
that re-energized the external shock to produce the P4 rebrightening.
The cause of the lack of $\gamma$-ray emission associated
with the delayed energy release is not clear and, lacking WFC data for
the P4 episode, it is difficult to constrain observationally;
though the slight count excess in the Konus soft band could be an
hint.
The delayed energy release may however be associated to the recycling of the
energy wasted while the relativistic jet propagates into the host star
(\cite{Mesz01, Ram02}). In that case, the
acceleration of the delayed fireball takes place at the star surface,
and is therefore characterized by a variability timescale many orders
of magnitude larger than that of the jet, effectively preventing the
occurrence of internal shocks.

We also note that among
the other GRBs afterglows for which the extrapolation of the decay law
to the prompt emission is inconsistent with the observed flux,
GRB 990704 (the X-ray richest event observed by \sax) is the only
analogous case. The afterglow X-ray flux decay of XRF 031203 also
shows an extrapolation below the ``probable'' prompt flux (\cite{Watson04}).
GRB 990510, 010222, 010214 show an extrapolation
\emph{above} the prompt emission, which is explained with a break a
few hours after the onset (\cite{Pian01, Zand01, Guidorzi03})

The peak width dependence as function of the energy was tested for P1
and P3 (see Fig.~\ref{f1_promptlc}).
To this aim we produced rebinned light curves
with bin size between 1 and 8 s. Their FWHM were obtained using
Gaussian fits; a 20\% systematic error on their estimate was added. By
using a law ${\rm FWHM}=kE^\alpha$ (expected by the synchrotron model,
Fenimore et al. (1995)) for the two peaks we obtain
$\alpha=-0.48\pm0.20$ for the P1 and $\alpha=-0.44\pm0.12$ for P3
(errors are 90\% confidence level). These results are consistent with
the results from the BATSE GRBs (\cite{Fenimore95}) as well as for
GRB 960720 (\cite{Piro98}) and 990704 (\cite{Feroci01}).

Due to the paucity of data and the complexity of the optical
light curve, it is not possible to constrain the fireball and
environment properties completely. From X-ray spectroscopy we infer
the electron distribution slope $p=2.1\pm0.25$ under the assumption
that X-rays are above the synchrotron cooling frequency. Due to the
large uncertainty, the X-ray decay slope of $\delta_{\rm X}=0.81\pm0.07$ can
be accounted for both in an ISM and wind environment. It is
tantalizing to note, however, that the early time optical slope seems
to be larger than the X-ray one. This would fit in a wind environment
scenario, consistent with the possible detection of a supernova bump
at late time (\cite{Levan04}). In this case one would expect
$\delta_{\rm X}=(3p-2)/4=1.1\pm0.2$ and
$\delta_{\rm O}=(3p-1)/4=1.3\pm0.2$, fully
consistent with the X-ray slope and the optical lower limit. Even this
interpretation bears however some degree of uncertainty. \grb\ has a
flux of 10.5 $\mu$Jy in R and $7.9\times 10^{-12}$ \fu\ in X-rays
which falls outside the distribution found by
\cite{depas02}) (see their Fig.~5) and would classify it as a dark
GRB (\cite{laz02}). Even assuming that the cooling frequency
lies exactly at the edge of the \sax\ band, the synchrotron spectrum
would over-predict optical emission by a factor $\sim5$. There are two
possible interpretations for this. One possibility is that the X-ray
emission is boosted by an IC component, like in the case of GRB 000926
(\cite{har01}). This would require a moderately dense environment,
either uniform or stratified. Alternatively, the optical emission may
be extincted by a sizable amount of dust in the host galaxy, with
$A_{\rm V}\sim2$. This would correspond, for a Galactic mixture, to a column
density $\nh\sim3\times10^{21}$~cm$^{-2}$, consistent with the upper
limit derived from X-ray spectroscopy. The lack of constraints on the
optical spectrum prevents us to reach a definite conclusion. The
optical spectrum should be bluer in the case of IC emission and red in
the case of dust obscuration.

Assuming that the emission line in the MECS spectra is real and due to
fluorescence of H-like iron (rest energy of 6.97 keV), then the change
in line position can be explained by a variable iron recombination
edge showing its maximum in the second half of ToO~1 (or later, but
before ToO~2). In fact if we derive the redshift from the line
position in ToO~2, we obtain $z\simeq1.7$ which leads to a
recombination edge of $\sim3.4$ keV.
Also the ratio between the iron recombination edge rest
energy, 9.28 keV, and 6.97 keV is $\simeq 1.3$, like for the ratio of
the ToO~1b over ToO~2 line energies.
Again, our statistics does not allow us to perform a simultaneous fit
for a Gaussian and a recombination edge line. However this hypothesis
appears in agreement with the data.

As no direct $z$ measurement exists for \grb, we calculated the peak
energy $E_{\rm p}$ in the $\nu F_\nu$ spectrum and the isotropic
energy $E_{\rm rad}$ for a grid of $z$ values. We then compared the
results with the relation reported by Amati et al. (2002). We find
that the relation is satisfied (with a discrepancy level $<20\%$) for
$0.9 < z < 1.5$ and $1.1\times10^{53}< E_{\rm rad} < 3.0\times10^{53}$
erg. This range of $z$ would exclude the value of 0.5 obtainable by
assuming a 1998bw-like SN re-bump (\cite{Levan04}) and is marginally in
agreement with the value derived above. Even assuming that our flux
estimate for the missing part of the X-ray light curve must be
increased by an extra 20\% (which is unlikely), the lower limit for $z$ 
becomes 0.6.
Besides this, we note that $z\simeq0.5$ together with the reported
magnitude of ${\rm V}\simeq 28.7$ for the host galaxy (\cite{Levan04})
would place it in the very low end of the galaxy luminosity function
(${\rm M_{\rm V}} = -14.3$),
which is unusual for GRB hosts; this independently of considering
the X-ray spectrum derived $\nh\simlt3\times10^{21}$~cm$^{-2}$ as being
``local'' or ``global''.

\begin{acknowledgements}
This research was supported by the Italian Space Agency (ASI) and
Consiglio Nazionale delle Ricerche (CNR). \sax\ was a major program of
ASI with participation of the Netherlands Agency for Aerospace
Programs (NIVR). All authors warmly thank the extraordinary teams of
the \sax\ Scientific Operation Center and Operation Control Center for
their enthusiastic support to the GRB program. One of of us (AJCT) is
grateful to the MOA project for granting part of the observing time to
the GRB follow-up program and thanks P. Kilmartin, A. Gilmore, Ph. Yock
and B. Nelson for their assistance.
This research made use of observations retrieved from the ESO data archive.
\end{acknowledgements}


\begin{thebibliography}{}
\bibitem[Amati et al. 2002]{Amati02}
Amati, L., Frontera, F., Tavani, M., et al. 2002, A\&A, 390, 81
\bibitem[Aptekar et al. 1995]{aptekar95}
Aptekar, R. L., Frederiks, D. D., Golenetskii, S. V., et al. 1995,
 Space Sci. Rev., 71, 265
\bibitem[Barraud et al. 2003]{barraud03}
Barraud, C., Olive, J-F., Lestrade, J. P., et al. 2003, A\&A, 400, 1021
\bibitem[Bj\"ornsson et al. 2002]{bj02}
Bj{\"o}rnsson, G., Hjorth, J., Pedersen, K., 
\& Fynbo, J. P. U. 2002, ApJ, 579, L59
\bibitem[Castro-Tirado et al. 1994]{ct94}
Castro-Tirado, A., Brandt, S., Lund, N., Lapshov, I. Y., Terekhov, O.,
 \& Sunyaev, R. A. 1994, in Gamma-ray Burst Workshop,
 ed. G. J. Fishman, J. J., Brainerd \& K. Hurley, AIP 307, 17
\bibitem[Castro-Tirado et al. 2001]{Alberto01}
Castro-Tirado, A., Sokolov, V. V., Gorosabel, J., et al. 2001, A\&A 370, 398
\bibitem[Castro-Tirado et al. 2002]{Albertonz02}
Castro-Tirado, A., Gorosabel, J., Castro Ceron, J. M., Nelson, B.,
 \& Tristram, P. 2002, GCN Circ., No. 1355
\bibitem[Costa 1999]{costa99}
Costa, E. 1999, A\&AS, 138, 425
\bibitem[Crew et al. 2003]{crew03}
Crew, G. B., Lamb, D. Q., Ricker, G. R., et al. 2003, ApJ, 599, 387
\bibitem[De Pasquale et al. 2002]{depas02}
De Pasquale, M., Piro, L., Perna, R., et al. 2002, ApJ, 592, 1018
\bibitem[Dermer \ 1999]{Dermer99}
Dermer, C. D., Chang, J. \& Bottcher, M. 1999, ApJ, 513, 656
\bibitem[Fenimore et al. 1995]{Fenimore95}
Fenimore, E. E., in 't Zand, J. J. M., Norris, J. P., Bonnell, J. T.,
 \& Nemiroff, R. J. 1995, ApJ, 448, L101
\bibitem[Feroci et al. 2001]{Feroci01}
Feroci, M., Antonelli, L. A., Soffitta, P., et al. 2001, A\&A, 378, 441
\bibitem[Frail et al. 2002]{frail02}
Frail, D., Wieringa, M. H., Berger, E., \& Wark, R. 2002, GCN Circ., No. 1380
\bibitem[Frontera et al. 2000]{Frontera00}
Frontera, F., Amati, L., Costa, E., et al. 2000, ApJS, 127, 59
\bibitem[Fruchter et al. 2002]{Fruchter02}
Fruchter, A., et al. 2002, GCN Circ., No. 1453
\bibitem[Fynbo et al. 2004]{Fynbo04}
Fynbo, J. P. U., Sollerman, J., Hjorth, J., et al. 2004, ApJ,
 accepted (astro-ph/0402240)
\bibitem[Gandolfi 2002]{Gandolfi02}
Gandolfi, G. 2002, GCN Circ., No. 1349
\bibitem[Granot et al. 2002]{Granot02}
Granot, J., Panaitescu, A., Kumar, P., \& Woosley, S. E., et al. 1999,
 ApJ, 570, L61
\bibitem[Guidorzi et al. 2003]{Guidorzi03}
Guidorzi, C., Frontera, F., Montanari, E., et al. 2003, A\&A, 401, 491
\bibitem[Heise et al. 2001]{Heise01}
Heise, J., in 't Zand, J. J. M.,  Kippen, M., Woods, P. 2001,
in ESO Proc. Ser., Gamma-Ray Bursts in the Afterglow Era, ed. E. Costa,
 F. Frontera \& J. Hjorth, Berlin Heidelberg: Springer, p. 16
\bibitem[Huang et al. 2002]{Huang02}
Huang, Y. F., Dai, Z. G., \& Lu, T. 2002, MNRAS, 332, 735
\bibitem[Jager et al. 1997]{jager97}
Jager, R., Mels, W. A., Brinkman, A. C.,  et al. 1997, A\&AS, 125, 557
\bibitem[Harrison et al. 2001]{har01}
Harrison, F. A., Yost, S. A., Sari, R., et al. 2001, ApJ, 559, 123
\bibitem[Kilmartin et al. 2002]{Kilmartin02}
Kilmartin, P., \& Gilmore, A. 2002, GCN Circ., No. 1350
\bibitem[Lazzati, Covino \& Ghisellini 2002]{laz02}
Lazzati, D., Covino, S., \& Ghisellini, G. 2002, MNRAS, 330, 583
\bibitem[Levan et al. 2004]{Levan04}
Levan, A., Nugent, P., Fruchter, A., et al. 2004, ApJ, submitted,
 astro-ph/0403450
\bibitem[M\'esz\'aros \& Rees 2001]{Mesz01}
M\'esz\'aros, P., \& Rees, M. J. 2001, ApJ, 556, L37
\bibitem[Mochkovitch et al. 2003]{moc03}
Mochkovitch, R., Daigne, F., Barraud, C., \& Atteia 2003,
to be published
in Proc. ``GRBs in the afterglow era: 3rd Rome workshop (2002)",
eds. L. Piro, F. Frontera, N. Masetti, M Feroci, astro-ph/0303289
\bibitem[Nicastro et al. 2002]{me02}
Nicastro, L., Piro, L., Gandolfi, G., Feroci, M., Capalbi, M., Perri, M.,
 Heise, J., \& in't Zand, J. J. M. 2002, GCN Circ., No. 1374
\bibitem[Paciesas et al. 1999]{Paciesas99}
Paciesas, W. S., Meegan, C. A., Pendleton, G. N., et al. 1999, ApJS, 122, 465
\bibitem[Panaitescu, M\'esz\'aros \& Rees 1998]{Panaitescu98}
Panaitescu, A., M\'esz\'aros, P., \& Rees, M. J. 1998, ApJ, 503, 314
\bibitem[Perri \& Capalbi 2002]{pc02}
Perri, M., \& Capalbi, M. 2002, A\&A, 396, 753
\bibitem[Pian et al. 2001]{Pian01}
Pian, E., Soffitta, P., Alessi, A., et al. 2001, A\&A, 372, 456
\bibitem[Piro et al. 1998]{Piro98}
Piro, L., Amati, L., Antonelli, A., et al. 1998, A\&A, 331, L41
\bibitem[Preece et al. 2002]{Preece00}
Preece, R. D., Briggs, M. S., Mallozzi, R. S., Pendleton, G. N.,
 Paciesas, W. S., \& Band, D. L. 2000, PpJS, 126, 19
\bibitem[Prochaska et al. 2003]{Prochaska03}
Prochaska, J., Bloom, J. S., Chen, H. W., Hurley, K., Dressler, A.,
 \& Osip, D. 2003, GCN Circ. 2482
\bibitem[Ramirez-Ruiz, Celotti \& Rees 2002]{Ram02}
Ramirez-Ruiz, E., Celotti, A., Rees, M. J. 2002, MNRAS, 337, 1349
\bibitem[Sari \& Piran 1999]{SaPi99} Sari, R. \& Piran, T., 1999, ApJ, 520, 641
\bibitem[$\check{\rm S}$imon et al. 2001]{simon01}
$\check{\rm S}$imon, V., Hudec, R., Pizzichini, G., \& Masetti, N. 2001, A\&A,
 377, 450 
\bibitem[Soderberg et al. 2003]{Soderberg03}
Soderberg, A. M., Kulkarni, S. R., Berger, E., et al. 2003, ApJ,
 submitted (astro-ph/0311050)
\bibitem[Yamazaki et al. \ 2002]{Yamazaki02}
Yamazaki, R., Yoka, K. S., \& Nakamura, T. 2002, ApJ, 571, L31
\bibitem[Watson et al. 2004]{Watson04}
Watson, D., Hjorth, J., Levan, A., et al. 2004, ApJ, accepted, astro-ph/0401225
\bibitem[Woosley 1993]{Woosley93}
Woosley, S. E. 1993, ApJ 405, 273
\bibitem[in 't Zand et al. 2001]{Zand01}
in 't Zand, J. J. M., Kuiper, L., Amati, L., et al. 2001, ApJ, 559, 710
\bibitem[in 't Zand et al. 2003]{Zand03}
in 't Zand, J. J. M., Heise, J., Kippen, R. M., et al. 2003, to be published
in Proc. ``GRBs in the afterglow era: 3rd Rome workshop (2002)",
eds. L. Piro, F. Frontera, N. Masetti, M Feroci, astro-ph/0305361
%
\end{thebibliography}
\end{document}